\documentclass[%
    reprint,
    10pt,
    superscriptaddress,
    longbibliography,
    nofootinbib,
    aps,
    prl
]{revtex4-2}
\usepackage{graphicx}

\usepackage{hyperref}
\hypersetup{
	colorlinks=true,
	linkcolor=blue,
	citecolor=blue,
	urlcolor=blue,
}

\usepackage{physics}
\usepackage{amsmath}
\usepackage{amssymb}
\usepackage{amsthm}
\usepackage{bm} 
\usepackage{bbm} 
\usepackage{mathtools}
\usepackage{color}
\usepackage[normalem]{ulem}

\usepackage{orcidlink}

\newcommand{\N}{\mathbb{N}}

\newcommand{\R}{\mathbb{R}}
\newcommand{\C}{\mathbb{C}}

\renewcommand{\d}{\mathrm{d}}

\newcommand{\Id}{\mathbbm{1}}

\renewcommand{\ket}[1]{|#1\rangle}
\renewcommand{\bra}[1]{\langle#1|}
\newcommand{\ketn}{\ket{\hspace{0.11em}\raisebox{-.5pt}{0}\hspace{0.10em}}}
\newcommand{\bran}{\bra{\hspace{0.10em}\raisebox{-.5pt}{0}\hspace{0.11em}}}
\newcommand{\nsp}{\hspace{-0.075em}}

\newcommand{\E}{\mathbb{E}}
\newcommand{\Tw}{\mathcal{T}_W}
\newcommand{\cG}{\mathcal{G}}
\newcommand{\Wcl}{W_{\hspace{-0.05em}\mathrm{cl}}}
\newcommand{\Wloc}{W_{\mathrm{loc}}}
\newcommand{\Wc}{W_{\hspace{-0.1em}c}}
\newcommand{\Wo}{W_{\hspace{-0.05em}0}}
\newcommand{\Wcval}{7.13}
\newcommand{\Wlocval}{67.0}
\newcommand{\rar}{{\rightarrow}}
\newcommand{\lar}{{\leftarrow}}
\newtheorem{theorem}{Theorem}
\newtheorem{lemma}{Lemma}

\begin{document}
\title{Localization transition in a spin chain with an exact Fock-space tree}

\author{Bernhard Jobst\,\orcidlink{0000-0001-7027-3918}}
\affiliation{Technical University of Munich, TUM School of Natural Sciences, Physics Department, 85748 Garching, Germany}
\affiliation{Munich Center for Quantum Science and Technology (MCQST), Schellingstr.~4, 80799 M\"unchen, Germany}
\affiliation{Center for Quantum Phenomena, Department of Physics, New York University, 726 Broadway, New York, New York 10003, USA}

\author{Simone Warzel\,\orcidlink{0000-0002-6250-4664}}
\affiliation{Technical University of Munich, TUM School of Computation, Information and Technology, Department of Mathematics, 85748 Garching, Germany}
\affiliation{Munich Center for Quantum Science and Technology (MCQST), Schellingstr.~4, 80799 M\"unchen, Germany}

\author{Alexander Altland\,\orcidlink{0000-0002-2991-4805}}
\affiliation{Institut f\"ur Theoretische Physik, Universit\"at zu K\"oln, Z\"ulpicher Stra{\ss}e 77, 50937 K\"oln, Germany}

\author{Frank Pollmann\,\orcidlink{0000-0003-0320-9304}\,}
\affiliation{Technical University of Munich, TUM School of Natural Sciences, Physics Department, 85748 Garching, Germany}
\affiliation{Munich Center for Quantum Science and Technology (MCQST), Schellingstr.~4, 80799 M\"unchen, Germany}

\begin{abstract}
Many-body localization is often pictured as Anderson localization on the graph of many-body configurations. Strongly correlated disorder and a dense net of loops, however, make this graph hard to analyze. We introduce a local disordered spin-1 chain in which a single domain wall explores an exact binary tree of configurations. Random fields create strongly correlated disorder on this tree, as in generic interacting systems, while the absence of loops keeps the problem tractable. We rigorously bound the spreading of the domain wall at strong disorder, and efficient numerics on exponentially large trees indicate a localization transition at finite disorder strength.
\end{abstract}

\maketitle

\textit{Introduction.}---Many-body localization (MBL)~\cite{BAA2006, GMP2005, NandkishoreHuse2015, Abanin2019} is an Anderson localization problem~\cite{Anderson1958} on a graph of many-body configurations, connected by transitions generated by the Hamiltonian~\cite{AGKL1997, LoganWelsh2019}. In microscopic systems, this Fock-space graph combines an exponentially large configuration space with strongly correlated diagonal energies and a complicated structure of loops~\cite{RoyLogan2020PRB, RoyLogan2020PRL, RoyLogan2024}. In view of this complexity, tree approximations have become a widely used route to analytic progress~\cite{AGKL1997, AbouChacra1973, MirlinFyodorov1991, TikhonovMirlin2021}. They retain the rapid growth of configuration space but omit the loops of the microscopic graph~\cite{RoyLogan2024}. This raises the question of whether a tree description can arise exactly from a local many-body Hamiltonian, while retaining the correlated disorder intrinsic to its Fock-space representation. Here we construct a disordered spin-1 chain whose dynamics decomposes into independent problems on exact binary trees, providing a microscopic setting for studying localization through rigorous bounds and efficient numerical evaluation of the cavity recursion.

\begin{figure}[ht]
    \includegraphics[width=\linewidth]{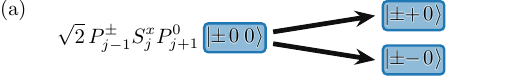}\\[2mm]
    \includegraphics[width=\linewidth]{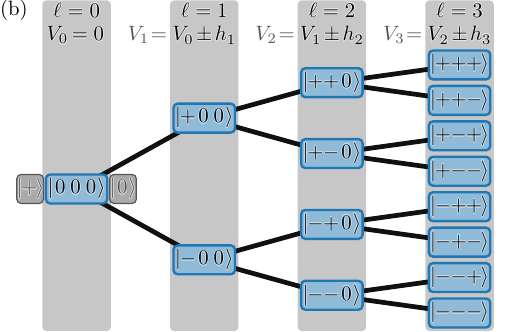}\\[2mm]
    \includegraphics[width=\linewidth]{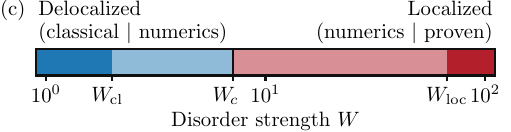}
    
    \caption{\label{fig:overview}%
    \textbf{Fock-space tree of a constrained spin-1 chain.} (a) The spin operator $\smash{\sqrt{2} \, P^\pm_{j-1}S^x_jP^0_{j+1}}$ flips the leftmost $\ketn$ into $\ket{\pm}$ (and back) with unit amplitude. (b) The configurations reachable from $\ket{000}$ (with fixed boundary spins shown in gray) form a binary tree; generation $\ell$ contains all $2^\ell$ sign strings of length $\ell$. The random fields assign the potential $V(\sigma) = \smash{\sum_j \sigma_j h_j}$, so that siblings differ by $2h_{\ell}$. (c) Localization bounds are proven for $W>\Wloc\approx\Wlocval$ (dark red), and the cavity recursion locates its onset numerically at $\Wc\approx\Wcval$ (light red; Fig.~\ref{fig:dynamics}). Below $\Wc$, the numerics indicate delocalization (light blue), and for $W<\Wcl=2$ resonant branches proliferate on average already in the classical limit (dark blue; End Matter~\protect\hyperlink{end:D}{D}).}
\end{figure}

Without disorder, the dynamics reduces to that of a freely propagating domain wall, which provides an analytically solvable reference point. Introducing randomness through local magnetic fields generates correlated disorder in Fock space, with two competing effects: neighboring configurations differ only by a local field contribution, favoring resonant propagation, while these differences accumulate along branches into a random-walk potential of increasing spread, opposing propagation. For this correlated potential, we establish exponential decay of energy-averaged Green's-function moments above a finite disorder threshold, uniformly in system size. We show how this implies bounds on the spreading of the previously freely propagating domain wall. A cavity recursion for each disorder realization further gives access to the survival probability and a localization criterion. Its numerical evaluation yields a critical disorder strength $\Wc\approx\Wcval$, indicating a localization transition at finite disorder in the thermodynamic limit, as summarized in the phase diagram of Fig.~\ref{fig:overview}c.

\textit{Model and Fock-space tree.}---Consider $N$ spin-$1$ degrees of freedom with a local basis $\{\ket{+},\ketn,\ket{-}\}$ governed by the Hamiltonian
\begin{equation}
    H = \sqrt{2} \sum_{j=1}^{N} P^{\pm}_{j-1} S^x_j P^0_{j+1} + \sum_{j=1}^{N} h_j S^z_j,
    \label{eq:H}
\end{equation}
where $P^0=\ketn\!\bran$, $P^\pm=\Id-P^0$ and $S^{x,z}$ are spin-1 operators, and the random field amplitudes $h_j$ are independent and uniformly distributed on $[-W,W]$. The prefactor $\sqrt{2}$ effectively introduces units in which the allowed spin flips have unit amplitude. The boundary terms are fixed by fictitious spins in the state $\ket{+}$ at $j=0$ and $\ketn$ at $j=N+1$. The kinetic term flips spin $j$ between the state $\ketn$ and the states $\ket{\pm}$ with unit amplitude, but only if its left neighbor is not in the state $\ketn$ and its right neighbor is. Starting from the all-zero state $\ketn^{\otimes N}\!$, every accessible configuration consists of a string of $\ket{+}$ and $\ket{-}$ spins followed by zeros. The boundary between these two regions defines a domain wall. Only spins adjacent to this wall can flip: converting the first zero into either $\ket{+}$ or $\ket{-}$ moves the wall one site to the right, while converting the last nonzero spin back into $\ketn$ moves it one site to the left, see Fig.~\ref{fig:overview}a. The states reachable from the all-zero state are thus labeled by sign strings $\sigma=(\sigma_1,\dots,\sigma_{\ell})$ with $\sigma_j=\pm1$ and $0\le\ell\le N$, with the convention $\sigma=0$ for $\ell=0$. Their connectivity graph is a binary tree rooted at $ 0 $ with $N+1$ generations and $2^{N+1}-1$ vertices, as shown in Fig.~\ref{fig:overview}b. The remaining product states form other Krylov sectors, each containing domain walls corresponding to smaller trees separated by frozen segments; the model is strongly fragmented~\cite{Sala2020, Khemani2020, Moudgalya2022}. We restrict attention to the largest sector, which comprises a fraction of order $(2/3)^N$ of the Hilbert space; a typical product state instead belongs to a sector of many small trees, on which the walls move independently and generically start from interior vertices rather than from the root. We expect the localization mechanism studied here to carry over to these cases but do not analyze them.

\textit{Dynamics of the clean system.}---Before turning to the localizing effects of disorder, we briefly discuss the ballistic domain-wall motion of the clean system. For $W=0$, symmetry under exchange of sibling subtrees~\cite{Childs2002, Kempe2003} confines the evolution from the root to uniform superpositions over configurations with the same domain-wall position $\ell$, where the first $\ell$ spins are nonzero and the remaining spins are zero. These configurations form generation $\ell$ of the Fock-space tree:
\begin{equation}
    \ket{\ell}
    = \ket{\rar}^{\otimes \ell}\otimes\ketn^{\otimes (N-\ell)}\nsp,
    \label{eq:layer_states}
\end{equation}
where $\ket{\rar}=(\ket{+}+\ket{-})/\sqrt2$. Thus $\ket{\ell}$ describes a domain wall at the bond between sites $\ell$ and $\ell+1$. Since $H\ket{\ell}=\sqrt2\,(\ket{\ell+1}+\ket{\ell-1})$, with the terms $\ket{-1}$ and $\ket{N+1}$ set to zero, the domain wall moves as a single particle on an open chain with hopping amplitude $\sqrt2$. In the infinite-chain limit it spreads ballistically with front velocity $v=2\sqrt2$, and the return probability is
\begin{equation}
    P_0(t) = \left[\frac{2\,J_1(vt)}{vt}\right]^2\!,
    \label{eq:clean_return}
\end{equation}
where $J_1$ is the Bessel function of the first kind; the long-time envelope decays as $t^{-3}$. The derivation, the full spectrum~\cite{AllardFroese2000, AizenmanWarzel2006}, and an entanglement bound are given in End Matter~\hyperlink{end:A}{A}.

\textit{Correlated disorder on the Fock-space tree.}---Turning to the effects of randomness, the local
fields $\{h_j\}$ are diagonal in the configuration basis and assign to the vertex $\sigma$ the potential
\begin{equation}
    V(\sigma)=\sum_{j=1}^{\ell}\sigma_j h_j.
    \label{eq:potential}
\end{equation}
Along each branch, $V$ performs a random walk whose $\ell$th step $\pm h_{\ell}$ is shared by the whole generation: siblings differ by $2h_{\ell}$, a vertex differs from its parent by at most $W\!$, and all $2^\ell$ sign combinations of the same $\ell$ fields occur in generation $\ell$. This structure of the Fock-space potentials differs from the independent potentials of the Anderson model on the Bethe lattice: the potentials are strongly correlated and unbounded, with typical magnitude $W\sqrt{\ell/3}$ in generation $\ell$. The disorder breaks the spin-exchange symmetry generated by $Q = \ket{+}\!\bra{-} + \ket{-}\!\bra{+} + \ketn\!\bran$ and destroys the clean solution, but $H$ retains the chiral symmetry $\{H,\mathcal{C}\}=0$ with $\mathcal{C}=\prod_j Q_j R_j$ and $R_j=e^{-i\pi S^z_j}$, which makes the spectrum symmetric about zero. As the sector dimension is odd, every realization has an exact zero mode~\cite{JonayPollmann2025}.

\textit{Localization at strong disorder.}---At sufficiently strong disorder, we rigorously bound the spreading of the domain wall launched from the all-zero state. Let $P_{\ell}(t)=\sum_{|\sigma|=\ell}|\bra{\sigma}e^{-iHt}\ket{0}|^2$ denote the probability of finding the wall at position $\ell$, and define its average over an observation time of order $T$ by
\begin{equation}
    \bar{P}_{\ell}(T)=\frac{2}{T}\int_0^\infty\hspace{-0.5em}\d t \,
    e^{-2t/T}P_{\ell}(t).
    \label{eq:time_average}
\end{equation}
For $W>\Wloc\approx\Wlocval$, a simple fractional moment bound  (see End Matter~\hyperlink{end:B}{B} for a proof) yields
\begin{equation}
    \E\bar{P}_{\ell}(T)
    \le C(W)(N+1) \, T^\alpha \, e^{-\ell/\xi(W)}\nsp,\qquad T\ge1,
    \label{eq:dynamical_bound}
\end{equation}
where $\E$ denotes the disorder average, $0<\alpha<1$, and $C(W)$ and $\xi(W)$ are finite and independent of $N$ and $T$; the factor $N+1$ arises because the width of the spectrum grows linearly with $N$. Optimizing for the slowest growth of the probability bound with observation time gives a limiting exponent $\alpha \approx 0.27$ as $W\searrow\Wloc$, with smaller exponents available at stronger disorder.

The bound decreases exponentially with distance and grows only as a sublinear power of observation time. Consequently, doubling the observation time is compensated by increasing the distance from the boundary by $\alpha\xi\ln 2$. This gives a confinement scale for the averaged domain-wall probability that grows at most logarithmically with $T$ and $N$, in contrast to the ballistic spreading of the clean system. This growth is a limitation of the simple bound: we do not expect it  to reflect the actual dynamics; the result does not establish confinement within a fixed distance for arbitrarily long times and system sizes, and $\Wloc$ is a sufficient threshold for the bound rather than the location of the transition.

The proof extends the fractional-moment method~\cite{AizenmanMolchanov1993, AizenmanWarzelBook} to the correlated potential. Its connection to the dynamics is supplied by Plancherel's identity,
\begin{equation}
    \bar{P}_{\ell}(T) =
    \frac{1}{\pi T}\int_{\mathbb R}\!\d E
    \sum_{|\sigma|=\ell}|G(0,\sigma;E+i/T)|^2,
    \label{eq:resolvent_time_average}
\end{equation}
where $G(\sigma,\sigma';z) = \bra{\sigma}(H-z)^{-1}\ket{\sigma'}$. On a tree, propagation from the root factorizes as
\begin{equation}
    G(0,\sigma;z) = (-1)^\ell \prod_{j=0}^{\ell} \Gamma_{\!\sigma^{(j)}}(z)
    \label{eq:factorization}
\end{equation}
into forward Green's functions along the unique branch of vertices $ \{ \sigma^{(j)} \} $ from the root $ (\sigma^{(0)} = 0 ) $ to the target $ (\sigma^{(\ell)} = \sigma ) $, cf.~\cite{Klein1998, AizenmanSimsWarzel2006}. The forward Green's functions are defined as $\Gamma_{\!\sigma}(z) = \bra{\sigma}(H_{\sigma}-z)^{-1}\ket{\sigma}$, where $H_{\sigma}$ is the restriction of $H$ to the subtree consisting of $\sigma$ and all its descendants, obtained by cutting the edge to its parent. They obey the recursion~\cite{AbouChacra1973}
\begin{equation}
    \Gamma_{\!\sigma}(z) =
    \Big[ V(\sigma) - z - \sum_{\sigma'} \Gamma_{\!\sigma'}(z) \Big]^{-1}\!,
    \label{eq:forward_recursion}
\end{equation}
where the sum runs over the children of $\sigma$. At the root, the subtree is the entire tree, so $\Gamma_{\!0}(z)=G(0,0;z)$.

For independent onsite disorder the factors in Eq.~\eqref{eq:factorization} can be controlled by successive averaging over the vertex potentials. In the present set-up, a single field also affects all later generations, so that procedure is unavailable. End Matter~\hyperlink{end:B}{B} shows how to conditionally extract randomness for a successive integration of fractional moments of the product in Eq.~\eqref{eq:factorization}. It proves the resulting fractional-moment bound and derives Eq.~\eqref{eq:dynamical_bound} from it.

\begin{figure}
    \includegraphics[
        width=\linewidth,
        trim={0.081cm 0.075cm 0.078cm 0.039cm},
        clip
    ]{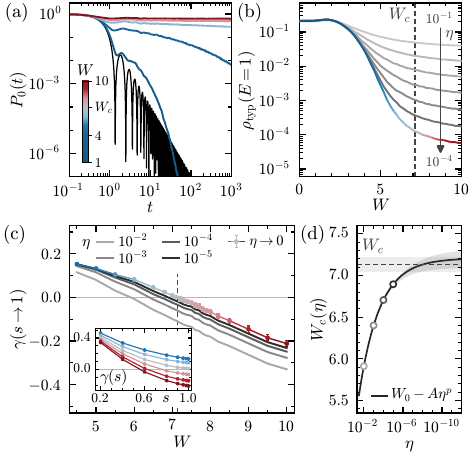}
    
    \caption{\label{fig:dynamics}%
    \textbf{Numerical cavity recursion.} (a) Survival probability of the all-zero state, $P_0(t) = \E |\bra{0}e^{-iHt}\ket{0}|^2$, from the Fourier transform of the local density of states of each realization ($N=400$ generations, $\eta=10^{-4}$, $10^3$ realizations) for $W=2,4,\ldots,10$ (color bar); the black line is the clean result $[2\,J_1(vt)/(vt)]^2$. (b) Typical local density of states $\rho_{\rm typ} = \exp(\E\ln\rho)$ of the all-zero state at the generic energy $E=1$ for broadenings from $\eta=10^{-1}$ to $10^{-4}$ ($N=10^3$, $10^4$ realizations). (c) Growth rate $\gamma(s=0.999)$ of the fractional-moment path sum generated by Eq.~\eqref{eq:qtransfer} versus $W$, extrapolated to the limit $\eta\to0$ (data points) and for $\eta=10^{-2}$ to $10^{-5}$ (gray; $N=400$, $50$ realizations); its sign change locates $\Wc=\Wcval\pm0.09$, marked by the dashed lines in (b)--(d). Inset: $\gamma(s)$ for $W=5,6,\ldots,10$, which decreases monotonically with $s$. (d) Finite-$\eta$ crossing points $\Wc(\eta)$ of (c); a fit $\Wo - A \eta^p$ extrapolates to $\Wo=7.21\pm0.11$, consistent with $\Wc$. Where not visible, statistical error bars are smaller than the plotted lines or symbols.}
\end{figure}

\textit{Cavity recursion.}---The correlated potentials also make the cavity method~\cite{AbouChacra1973} numerically inexpensive. For independent potentials, as on the Bethe lattice, the subtrees are statistically independent, and the recursion becomes a self-consistent equation for the distribution of the forward Green's functions $\Gamma_{\!\sigma}$ in Eq.~\eqref{eq:forward_recursion}, solved numerically by population dynamics. For the correlated potentials considered here, this approach neither applies nor is needed: all vertices of generation $\ell$ have identical forward subtrees up to the offset $V(\sigma)$ of their root, so that 
\begin{equation}
    \Gamma_{\!\sigma}(z) = \cG_{\ell}\big(z-V(\sigma)\big)
    \label{eq:singleF}
\end{equation} 
with a single function per generation. Hence Eq.~\eqref{eq:forward_recursion} turns into
\begin{equation}
    \cG_{\ell}(z) = - \Big[
    z + \cG_{\ell+1}(z-h_{\ell+1}) + \cG_{\ell+1}(z+h_{\ell+1})
    \Big]^{-1}\!,
    \label{eq:cavity}
\end{equation}
with $\cG_N(z) = -1/z$ at the leaves. For a given realization $\{h_j\}$, Eq.~\eqref{eq:cavity} is iterated from the leaves to the root on an energy grid, with interpolation off the grid. It yields the local density of states $\rho(E) = \pi^{-1}\,{\rm Im}\,\cG_0(E+i\eta)$ of the all-zero state at a cost linear in $N$ (End Matter~\hyperlink{end:C}{C}).

\textit{Survival probability.}---The most direct observable is the probability $P_0(t)=|\bra{0}e^{-iHt}\ket{0}|^2 = |\!\int\!\d E \, \rho(E)\,e^{-iEt}|^2$ that the chain is still in the all-zero state, shown in Fig.~\ref{fig:dynamics}a after averaging over realizations. For $W=0$ it follows Eq.~\eqref{eq:clean_return}, with a $t^{-3}$ decay envelope. Weak disorder leaves the decay to zero intact but slows it down; beyond $\Wc$ (determined below) the average $\E P_0$ saturates, on the accessible time scale, at a finite plateau, whose long-time average is the inverse participation ratio of the all-zero state in the eigenbasis, and the chain retains memory of its initial state. Just below $\Wc$ a very slow decay is visible on the accessible time scale. The conventional diagnostic of localization, the typical local density of states at a generic energy, $\exp(\E\ln\rho)$, is shown in Fig.~\ref{fig:dynamics}b: it is independent of $\eta$ and close to the clean value for $W\lesssim2$, and, beyond it, decreases with decreasing $\eta$, approaching the pure-point behavior $\rho_{\rm typ}\propto\eta$ at large $W$. As on the Bethe lattice~\cite{TikhonovMirlin2021}, however, the crossover between the two behaviors is broad at any accessible $\eta$ and does not by itself locate the transition, which is why we turn to a more direct criterion.

\textit{Fractional-moment criterion.}---Localization requires propagation away from the root to be suppressed despite the exponentially growing number of accessible configurations. To test this competition, we sum the magnitudes of the Green's functions connecting the root to all configurations in generation $\ell$, each raised to a certain power $0<s<1$. Exponential decay of this sum with $\ell$ provides a localization criterion~\cite{AbouChacra1973, AizenmanWarzel2013}. The cavity recursion in Eq.~\eqref{eq:cavity} allows us to evaluate this sum without enumerating the exponentially many configurations: combining Eqs.~\eqref{eq:factorization} and~\eqref{eq:singleF}, the sum over all $2^\ell$ branches, $\Phi_{\ell}(s) = \sum_{|\sigma|=\ell} \prod_{j=0}^{\ell} \big|\cG_j\big(E + i0^{+} - V(\sigma^{(j)})\big)\big|^{s}\!$, is generated by the transfer recursion 
\begin{equation}
    \varphi_j(V) = \big|\cG_j(E-V)\big|^{s} \,\nsp \big[\varphi_{j\nsp-\nsp1}(V\nsp\nsp- h_j) + \varphi_{j\nsp-\nsp1}(V\nsp\nsp+ h_j)\big],
    \label{eq:qtransfer}
\end{equation}
with $\varphi_0(V)=|\cG_0(E)|^{s} \, \delta(V)$ and $\Phi_{\ell} = \int\!\d V \, \varphi_{\ell}(V)$ (End Matter~\hyperlink{end:C}{C}).
The competition between decay along individual branches and the proliferation of branches is measured by the growth rate $\gamma(s) = \lim_{\ell\to\infty} \ell^{-1} \ln\Phi_{\ell}(s)$. Localization is indicated when the summed propagation amplitudes decay exponentially, corresponding to $\gamma(s)<0$ for some $0<s<1$~\cite{AbouChacra1973, AizenmanWarzel2013}. We take the sign change of the growth rate $\inf_{0<s<1}\gamma(s)$, as our estimate of the transition. For independent potentials on regular trees, the sign of this growth rate as $s\to1$ separates localized from delocalized spectral regimes~\cite{AizenmanWarzel2013}; for the correlated potentials considered here, a negative growth rate signals the exponential decay of fractional moments that underlies the dynamical bound of End Matter~\hyperlink{end:B}{B}, consistent with the survival probability and local density of states in Fig.~\ref{fig:dynamics}.

Two features are specific to the model. First, when Eq.~\eqref{eq:qtransfer} is written in the variable $E-V$, the energy $E$ enters the transfer recursion Eq.~\eqref{eq:qtransfer} only through the initial condition (End Matter~\hyperlink{end:C}{C}), so that $\gamma(s)$, and with it $\Wc$, is expected to be the same at every energy: we do not expect a mobility edge in energy because the random walk of the potential visits every energy along the tree. Second, Eq.~\eqref{eq:qtransfer} is evaluated for each realization, rather than through the distribution of the forward Green's functions as on the Bethe lattice, and we find $\gamma(s)$ to be self-averaging, with sample-to-sample fluctuations of order $10^{-2}$ (End Matter~\hyperlink{end:C}{C}). Theorem~\ref{thm:main} bounds the growth rate of the energy- and disorder-averaged path sum by $\ln r_s(W)$, with $r_s(W)$ defined in Eq.~\eqref{eq:decay_length}; the growth rate is thus negative for $W>\Wloc$. The inset of Fig.~\ref{fig:dynamics}c shows that $\gamma(s)$ decreases monotonically with $s$, so that the condition is first met as $s\to1$. In the limit $\eta\to0$, $\gamma(s\to1)$ changes sign at $\Wc\approx\Wcval$ (Fig.~\ref{fig:dynamics}c); a finite broadening shifts the crossing to smaller $W$, and extrapolating the finite-$\eta$ crossings gives a consistent value (Fig.~\ref{fig:dynamics}d and End Matter~\hyperlink{end:C}{C}). The value of $\Wc$ lies well inside the region where the typical local density of states of Fig.~\ref{fig:dynamics}b has already dropped by more than three orders of magnitude at $\eta=10^{-4}$, so that finite-$\eta$ diagnostics make the regime below $\Wc$ appear localized; there, according to the fractional-moment criterion, the all-zero state is delocalized but relaxes extremely slowly, consistent with the survival probability at $W=6$ in Fig.~\ref{fig:dynamics}a.

\textit{Resonant branches.}---The classical limit of Eq.~\eqref{eq:qtransfer} suggests an explanation for why delocalization survives the unbounded potential at all energies for sufficiently small $ W $. Replacing the resonance factor $|\cG_j(E-V)|^{s}$ by the indicator of the window $|V|\le1$ set by the hopping turns $\Phi_{\ell}$ into the number of branches along which the potential stays resonant up to generation $\ell$, a count that is exactly solvable (End Matter~\hyperlink{end:D}{D}): its disorder average grows as $\Lambda(W)^{\ell}$ with $\Lambda>1$ for $W<2$ and $\Lambda=2/W\le1$ for $W\ge2$. Resonant branches thus proliferate exponentially on average up to $W=2$ despite the $\sqrt{\ell}$ growth of the typical potential, and off-resonant tunneling, which the count discards, can extend delocalization up to $\Wc$~\cite{Anderson1958, AbouChacra1973, RoyChalkerLogan2019}.

\textit{Discussion and outlook.}---We have established a localization transition in a constrained spin-1 chain, with domain-wall propagation as the diagnostic observable. Rigorous bounds establish a dynamically localized regime above the conservative threshold $\Wloc\approx\Wlocval$, while the cavity recursion locates the transition numerically at $\Wc\approx\Wcval$ (Fig.~\ref{fig:overview}c), which we argued to be independent of energy. For a full mathematical proof of the localization--delocalization transition, it remains to establish the stability of the delocalized phase to weak disorder.
Of the three features that set MBL apart from single-particle Anderson localization---the exponential growth of Fock space, its strongly correlated disorder, and its complex structure of loops---our model retains the first two, albeit with a fixed rather than extensive number of neighbors, and removes the loops. This makes the problem tractable: on the tree, the Green's function factorizes along branches, which underlies the proof, and all vertices of a generation share one Green's function up to an energy shift, so that the cavity recursion can be evaluated at a cost linear in system size. While our analysis focuses on one of the largest trees, we expect the same localization mechanism to govern the other trees in the fragmented Fock space and hence to apply to generic initial states. The model thus provides a microscopic setting in which localization in Fock space is controlled rigorously at strong disorder.

Within this particular model, two directions of further exploration stand out. First, in the clean model the entanglement entropy stays bounded because the layer states of Eq.~\eqref{eq:layer_states} are products; disorder converts them into entangled superpositions, and it remains to be seen whether the entanglement then grows linearly below $\Wc$ and logarithmically above it, as in ergodic~\cite{KimHuse2013} and many-body localized chains~\cite{ZnidaricProsenPrelovsek2008, BardarsonPollmanMoore2012}. Second, even in the delocalized phase most branches leave the resonance window, so that eigenstates may occupy only a vanishing fraction of the tree; whether this leads to multifractal eigenstates, as discussed for the Anderson model on finite trees~\cite{TikhonovMirlin2021}, is open.

Natural extensions bring the model closer to generic MBL. Spin-$S$ chains in which the constrained flip connects $\ket{0}$ to all $2S$ other states realize $2S$-ary trees. They raise the coordination number, which is extensive in generic chains, and would allow testing microscopically the $\sqrt{2S}$ scaling predicted by a self-consistent theory for Cayley trees with correlated Gaussian potentials~\cite{RoyLogan2020PRL}. The more central question is how loops change this picture. In generic chains, loops arise because different regions of the chain can rearrange independently: performing two such rearrangements in either order leads to the same configuration and closes a loop. Here they are absent because a single domain wall carries all the dynamics. Letting the kinetic term also exchange $\ket{+}$ and $\ket{-}$ connects siblings and adds small, local loops, turning the tree into a tree of triangles, for which it may be possible to adapt the cavity recursion and the factorization along branches. In this sense, our model approaches generic MBL from the limit opposite to the quantum random energy model on the hypercube~\cite{Baldwin2016, ManaiWarzel2020, ManaiWarzel23}, which has an extensive number of neighbors and loops on all scales, but independent random energies without microscopic origin.

\begin{acknowledgments}
\textit{Acknowledgments.}---We acknowledge support from the Deutsche Forschungsgemeinschaft (DFG, German Research Foundation) under Germany's Excellence Strategy--EXC--2111--390814868, TRR 360--492547816, the DFG Research Unit FOR 5522 (project id 499180199), as well as the Munich Quantum Valley, which is supported by the Bavarian state government with funds from the Hightech Agenda Bayern Plus.
B.J.\ acknowledges support from the Gordon and Betty Moore Foundation through a postdoctoral fellowship at New York University, Grant No.~GBMF13921.
A.A.\ acknowledges partial support from the Deutsche Forschungsgemeinschaft (DFG) through the Cluster of Excellence Matter and Light for Quantum Computing (ML4Q), EXC 2004/1 (grant 390534769), and through the CRC network TR 183 (grant 277101999), project A03.
\end{acknowledgments}

\textit{Data availability.}---Data, data analysis, and simulation codes will be made available on Zenodo.

\textit{Use of AI tools.}---Claude Code (Opus 5.5, Fable 5.1) was used to assist with algebraic consistency checks, code development, literature review, and language editing. All AI-assisted material was reviewed and revised by the authors, who take full responsibility for the content.

\bibliography{bibliography}

\section*{End Matter}

\hypertarget{end:A}{\textit{A. Clean model.}}---The layer states of Eq.~\eqref{eq:layer_states} reduce the dynamics from the root to a hopping problem with amplitude $\sqrt2$. In the infinite system, the standing-wave eigenfunctions on the semi-infinite half-line are $\braket{\ell}{\psi_k}=\sqrt{2/\pi}\,\sin[k(\ell+1)]$, with dispersion $\epsilon(k) = 2\sqrt2\cos k = v \cos k$, $k\in(0,\pi)$. The maximal group velocity is $\max_k|\partial_k\epsilon(k)| = v = 2\sqrt2$. The propagator is
\begin{equation}
    \bra{\ell}e^{-iHt}\ket{0}
    = \frac{2}{\pi} \int_0^{\pi}\!\d k \, e^{-i\epsilon(k)t} \sin[k(\ell+1)]\sin(k).
\end{equation}
Bessel's integral and the recurrence $J_{\ell}(z)+J_{\ell+2}(z)=\frac{2(\ell+1)}{z}J_{\ell+1}(z)$~\cite{DLMF} give
\begin{equation}
    \bra{\ell} e^{-iHt}\ket{0}
    = (-i)^{\ell}\,\frac{2(\ell+1)}{vt}\,J_{\ell+1}(vt).
    \label{eq:bessel}
\end{equation}
Setting $\ell=0$ yields Eq.~\eqref{eq:clean_return}; the large-argument asymptotics of $J_1$ gives the $t^{-3}$ envelope. Every individual sign string of generation $\ell$ carries the amplitude $2^{-\ell/2}\bra{\ell}e^{-iHt}\ket{0}$.
On a finite tree with $N+1$ generations the standing waves $\braket{\ell}{\psi_n} \propto \sin[\pi n(\ell+1)/(N+2)]$, $n=1,\ldots, N+1$, diagonalize $H$ in the span of the layer states.

\emph{Full spectrum.} The full spectrum in the all-zero-state sector can be constructed making use of the state $\ket{\lar}=(\ket{+}-\ket{-})/\sqrt2$, which is annihilated by $S^x$ and hence acts as an impenetrable wall: placed at site $\ell+1$ behind an arbitrary sign string on the first $\ell$ sites, it decouples a chain of $L=N-\ell-1$ sites. The eigenvalues are therefore the standing-wave energies of the $(L+1)$-site hopping chains $\epsilon_{L+1}(n) = 2\sqrt2\cos[\pi n/(L+2)]$, $n=1,\dots,{L+1}$, corresponding to the smaller $L$-spin subsystems, with multiplicity $2^{\ell} = 2^{N-L-1}$ for $L=0,\dots,N-1$. These energies are complemented by the $N+1$ standing waves with energies $\epsilon_{N+1}(n)$ of the full tree.
The corresponding eigenstates, the standing waves of the full tree together with those on the subsystems that have the state $\ket{\lar}$ at site $\ell+1$, are mutually orthogonal: for equal $\ell$ they differ in the leading sign string or are distinct standing waves of the same chain, and otherwise the state $\ket{\lar}$ faces $\ket{\rar}$ or $\ket{0}$ in the other state. Since their number, $\sum_{L=0}^{N-1}(L+1)\,2^{N-L-1}+(N+1)=2^{N+1}-1$, equals the sector dimension, they form a complete eigenbasis (cf.\ Refs.~\cite{AllardFroese2000, AizenmanWarzel2006}).

\emph{Entanglement.} States reached during the evolution of the all-zero state satisfy an entanglement bound, which follows from the Schmidt decomposition across a cut between sites $\ell$ and $\ell+1$: all layer states $\ket{j}$ with $j\le\ell$ end in $\ket{0}^{\otimes(N-\ell)}$ and all with $j>\ell$ begin with $\ket{\rar}^{\otimes\ell}$. Grouping the terms of $\sum_j a_j(t)\ket{j}$ accordingly writes the state as a sum of two product states across the cut. Its Schmidt rank is therefore at most two, and the entanglement entropy never exceeds $\ln2$.

\hypertarget{end:B}{\textit{B. Localization bound and proof.}}---We first state the fractional-moment estimate underlying Eq.~\eqref{eq:dynamical_bound}, then prove it by successive averaging along a branch, and finally derive the dynamical consequence. We use the Green's function $G$ defined in the main text and write $\E$ for the disorder average.

\begin{theorem}\label{thm:main}
    Let $s\in(0,1)$ and $C_s=(8/\pi)^{s/2}(1-s)^{-1}$. For all $N$, all $\ell\le N$, all $\eta>0$, and every energy interval $I$ of length $2W$,
    \begin{equation}
    \begin{aligned}
        &%
        \frac{1}{|I|}\int_I\!\d E \sum_{|\sigma|=\ell}\E\,\big|G(0,\sigma;E+i\eta)\big|^{s}\\
        &\hspace{4em}%
        \le \sqrt{\tfrac{\pi e}{8}}\;(\ell+2)^{s/2}\bigg[\frac{\sqrt{2\pi e}\,C_s}{W^{s}}\bigg]^{\ell+1}\!\equiv A_\ell .
        \label{eq:fmb}
    \end{aligned}
    \end{equation}
\end{theorem}

\noindent For later use, define
\begin{equation}
    r_s(W) \equiv \frac{\sqrt{2\pi e}\,C_s}{W^s}
    \quad\text{and}\quad
    \xi_s^{-1} \equiv -\ln r_s(W).
    \label{eq:decay_length}
\end{equation}
The decay length $\xi_s$ is positive and finite whenever $r_s(W)<1$. Such an exponent
$s\in(0,1)$ exists for
\begin{equation}
    W>\Wloc\equiv\min_{0<s<1}
    (\sqrt{2\pi e}\,C_s)^{1/s}\approx\Wlocval,
    \label{eq:sstar}
\end{equation}
with the minimum attained at $s_*\approx0.73$. The bound in Theorem~\ref{thm:main} is uniform in the broadening $\eta$ and in $N$.

\emph{Proof strategy and energy averaging.} Key to the proof is the factorization of the Green's function $G(0,\sigma;z)$ along the path $\{ \sigma^{(j)} \}$ connecting the root and $\sigma = (\sigma_1, \dots , \sigma_{\ell})$, cf. Eq.~\eqref{eq:factorization}. Since the probability distribution of the onsite fields $h_j$ is symmetric under $h_j\to-h_j$, which flips $\sigma_j$, we may take each $\sigma^{(j)}=(+,\dots,+)$ without loss of generality. The proof will isolate in each factor of the product in Eq.~\eqref{eq:factorization} a random variable absent from all later factors. Its average yields a suppression factor proportional to $W^{-s}$. 

To make the $W\!$-dependence explicit, we write $h_j=W\tilde h_j$ with $\tilde h_j$ uniform on $[-1,1]$ and drop the tilde. Since the root has no potential to average over, we add a global random energy $h_0$ distributed uniformly on $[-1,1]$ and independent of the other fields; then $H'=H+Wh_0$ has $G'(z)=G(z-Wh_0)$, so averaging over $h_0$ is the energy average over an interval of length $2W$ in Eq.~\eqref{eq:fmb}. The potential of $H'$ is $W U(\sigma)$ with $U(\sigma)=h_0+\sum_{j\le\ell}\sigma_jh_j$. 
In the following, we abbreviate $ \Gamma_{\!j} = \Gamma_{\!\sigma^{(j)}}(z)$ and $\Gamma^{+}_{\!j}=\Gamma_{\!j+1}$ for $j<\ell\leq N$, and denote by $\Gamma_{\!j}^{-} $ the forward Green's function of the minus direction branching from the plus path, so that $$\Gamma_{\!j} = \big[ W U(\sigma^{(j)}) - z - \Gamma_{\!j}^{+} - \Gamma_{\!j}^{-} \big]^{-1}\!.$$ 

\emph{Gaussian rotation.} When $\ell=N$, we adjoin an auxiliary field $h_{N+1}$ that does not enter the Hamiltonian, so that the same rotations can be used at the last step. Assume first that $h_0,\dots,h_{\ell+1}$ are drawn from independent standard Gaussian distributions (the distribution of the onsite fields of later generations is irrelevant). Let $S_j=h_0+\dots+h_j$, $u_j=S_j/\sqrt{j+1}$, and define the orthogonal rotation
\begin{equation*}
    u_{j+1}=\frac{\sqrt{j+1}\,u_j+h_{j+1}}{\sqrt{j+2}}
    \hspace{0.5em}\text{and}\hspace{0.5em}
    v_j=\frac{u_j-\sqrt{j+1}\,h_{j+1}}{\sqrt{j+2}} ,
\end{equation*}
so that $v_j$ and $u_{j+1}$ are also sampled from independent standard Gaussians, independent of $h_{j+2},h_{j+3},\dots$ With $\alpha_j=\sqrt{(j+1)/(j+2)}$ one finds
\begin{equation}
\begin{aligned}
    &&U(\sigma^{(j)}) = S_j &= \alpha_j\big(\sqrt{j+1}\,u_{j+1}+v_j\big),\\
    &&S_j+h_{j+1} &= \sqrt{j+2}\,u_{j+1}\\
    &\text{and}&S_j-h_{j+1} &= \alpha_j\Big(\tfrac{j}{\sqrt{j+1}}\,u_{j+1}+2v_j\Big).
    \label{eq:rotated}
\end{aligned}
\end{equation}
For $j<N$, all potentials in the subtree of the $\pm$ child of $\sigma^{(j)}$ are $W(S_j \pm h_{j+1})$ plus contributions from $h_{j+2}, h_{j+3}, \ldots$
Consequently $v_j$ enters $\Gamma_{\!j}$ only through the onsite term $W\alpha_j v_j$ and through a uniform shift $2W\alpha_jv_j$ of all potentials of the $-$ subtree; i.e., $\Gamma^{-}_{\!j}(z)=\tilde\Gamma^{-}_{\!j}(z-2W\alpha_jv_j)$ with $\tilde\Gamma^{-}_{\!j}$ independent of $v_j$, while $\Gamma^{+}_{\!j}=\Gamma_{\!j+1}$ is independent of $v_j$, and all later $\Gamma_{\!k}$ with $k>j$ are functions of $(u_{j+1},h_{j+2},\dots)$ alone.

\emph{Averaging lemma.} The following standard bound handles fractional moments of $\Gamma_{\!j}$ averaged over $v_j$, which enters both the onsite term and the off-branch forward Green's function $\Gamma^{-}_{\!j}$ through a shift of its argument, cf.~\cite[Ch.~8]{AizenmanWarzelBook}.
\begin{lemma}\label{lem:avg}
    Let $z=E+i\eta$ with $\eta>0$, let $\Gamma^{\pm}$ be Stieltjes transforms of probability measures, and $D(v)=[- v+\kappa-z-\Gamma^{+}(z)-\Gamma^{-}(z+ c v)]^{-1} $ with $ b,c > 0 $, $\kappa\in\R$. Then for every $s\in(0,1)$ and every bounded probability density $\rho$,
    \begin{equation}
        \int_{\R}\!\d v \, \rho(v) \, |D(bv)|^{s}
        \le \frac{1}{1-s}\bigg(\frac{4}{b}\|\rho\|_\infty\bigg)^{\!s}\!.
        \label{eq:lemma}
    \end{equation}
    \end{lemma}
\begin{proof}
For any $\eta > 0$, the function $\zeta \mapsto D(\zeta)$ (i) is analytic on a neighborhood of $\C^{+} \cup \R $, (ii) maps $\C^{+}$ into itself, and (iii) obeys $D(iy)\to 0$ and $y\,{\rm Im}\,D(iy)\to1$ for $y\to\infty$ since $\Gamma^{-}(a+iy)\to0$ for any $a \in \R$. By the Herglotz representation theorem~\cite{AizenmanWarzelBook}, that function is the Stieltjes transform of a probability measure, with boundary values $D(v)$ on $\R$. Such functions obey the weak-$L^1$ bound $|\{w\in\R:| D(w)|\ge t\}|\le 4/t$~\cite[Prop.~8.2]{AizenmanWarzelBook}. The assertion then follows by scaling $ v \to b v $ from the layer-cake representation and a change of variables:
\begin{align*}
  \int_{\R}\! \d v \, \rho(v) \, |D(bv)|^{s}
  &= \int_0^\infty\hspace{-0.5em}\d t \,s\,t^{s-1}
     \!\int_{|D(w)|\ge t}\hspace{-1.5em}\d w \, \rho(w/b) / b \\
  &\le\int_0^\infty\hspace{-0.5em}\d t \,s\,t^{s-1}
     \min\Big\{1,\frac{4\|\rho\|_\infty}{b t}\Big\},
\end{align*}
which evaluates to Eq.~\eqref{eq:lemma}.
\end{proof}

\emph{Iteration.} We condition on $(u_{j+1},h_{j+2},\dots)$ and integrate over $v_j$: by Eqs.~\eqref{eq:factorization},~\eqref{eq:forward_recursion} and~\eqref{eq:rotated}, we have $\Gamma_{\!j}=D(-bv_j)$ with $D$ as in Lemma~\ref{lem:avg} with $b=W\alpha_j$, $\kappa=W\alpha_j\sqrt{j+1}\,u_{j+1}$, $c=2$, and $\Gamma^\pm$ independent of $v_j$, while $\prod_{k>j}|\Gamma_{\!k}|^s$ does not depend on $v_j$ at all. Applying the lemma with the standard Gaussian density $\rho=\rho_{\mathcal{N}}$, for which $\|\rho_{\mathcal{N}}\|_\infty=(2\pi)^{-1/2}$, successively for $j=0,1,\dots,\ell$ gives
\begin{equation}
    \E_{\mathcal N}\prod_{j=0}^{\ell}|\Gamma_{\!j}|^{s}\le\prod_{j=0}^{\ell}\frac{C_s}{(W\alpha_j)^{s}}=(\ell+2)^{s/2}\bigg(\frac{C_s}{W^{s}}\bigg)^{\!\nsp\ell+1}\!. 
\end{equation}
At a leaf, both child terms are absent; direct averaging of $|W\alpha_jv_j+\kappa-z|^{-s}$ gives the same bound (or one may repeat the averaging lemma with $\Gamma^\pm=0$). Here $C_s = (1-s)^{-1}(8/\pi)^{s/2}$ as in Theorem~\ref{thm:main}, and the product $\prod_{j=0}^{\ell}\alpha_j^{-2}=\ell+2$ telescopes.

\emph{Uniform distribution and summation.} The pointwise bound $\tfrac12\bm{1}_{[-1,+1]}(x)\le\sqrt{\pi e/2}\,\rho_{\mathcal{N}}(x)$, applied to the $\ell+2$ fields $h_0,\dots,h_{\ell+1}$, gives $\E_{\rm box}[\,\cdot\,] \le (\pi e/2)^{(\ell+2)/2} \E_{\mathcal N}[\,\cdot\,]$ for nonnegative functions of the fields. Multiplying by the number $2^\ell$ of vertices in generation $\ell$ yields Eq.~\eqref{eq:fmb}; the bound is uniform in $z$, in $N$, and in the choice of $\sigma^{(\ell)}$.

\emph{From fractional moments to dynamics.} In order to connect the results of Theorem~\ref{thm:main} to the dynamics of the domain wall, we apply the bound in Eq.~\eqref{eq:fmb} to Eq.~\eqref{eq:resolvent_time_average}.

We first keep $\eta>0$ arbitrary. The resolvent norm bound applied to the Green's function gives $|G|^2\le\eta^{s-2}\,|G|^s$. The operator's spectrum lies in $[-B,B]$, with $B=2\sqrt2+WN$. For $|E|>B+1$ we can use the stronger estimate $|G|^2\le(|E|-B)^{s-2}|G|^s$. Covering $[-B-1,\,B+1]$ takes $N+\lceil(2\sqrt{2}+1)/W\rceil=N+1$ intervals of length $2W$ since $W>\Wloc$ is much larger than $2\sqrt{2}+1$. This bounds the contribution of $[-B-1,\,B+1]$ to the energy integral in Eq.~\eqref{eq:resolvent_time_average} by $2W(N+1)\eta^{s-2}A_{\ell}$ with $A_{\ell}$ from the right-hand side of Eq.~\eqref{eq:fmb}. On each of the two tails, we use consecutive intervals of length $2W$ starting at distance one from the spectral edge. Their contributions are bounded by $2W A_{\ell}\sum_{k\ge0}(1+2Wk)^{s-2}$ per tail. The series converges because $s-2<-1$ and is bounded by a constant $\tilde{c}_s$ depending only on $s$ for $W>\Wloc>1$. Thus, for $0<\eta\le1$,
\begin{equation}
\begin{aligned}
    &\int_{\mathbb R}\!\d E\sum_{|\sigma|=\ell}
      \E|G(\sigma,0;E+i\eta)|^2\\
    &\hspace{4em}\le \tilde{c}_s W(N+1)\eta^{s-2}A_{\ell} .
\end{aligned}
\end{equation}
Finally, we set $\eta=T^{-1}\!$, $T\ge1$, and multiply by $1/(\pi T)$ as in Eq.~\eqref{eq:resolvent_time_average}. Since $W r_s(W)^{\ell+1} = \sqrt{2\pi e} \, C_s \, W^{1-s} \, e^{-\ell/\xi_s}$, this yields
\begin{equation}
    \E\bar{P}_{\ell}(T)\le
    c_s \, (N+1) \, (WT)^{1-s} \, (\ell+2)^{s/2} \, e^{-\ell/\xi_s}\!,
    \label{eq:dynamical_bound_detailed}
\end{equation}
after absorbing numerical factors into $c_s$. The conversion introduces $\eta^{s-1}=T^{1-s}$: uniformity of the fractional-moment bound in $\eta$ does not make this dynamical bound uniform in time.

To obtain the simpler form given in Eq.~\eqref{eq:dynamical_bound}, we choose any $s\in(0,1)$ satisfying $r_s(W)<1$ and set $\alpha=1-s$. Define the localization length $\xi(W)=2\xi_s(W)$ and the constant $M(W)=\sup_{j\in\N_0}(j+2)^{s/2}\,e^{-j/[2\xi_s(W)]}<\infty$. This gives $(\ell+2)^{s/2} \, e^{-\ell/\xi_s} \le M(W) \, e^{-\ell/\xi(W)}$. Taking $C(W) = c_s \, W^{1-s} \, M(W)$ proves Eq.~\eqref{eq:dynamical_bound}; the dependence of these constants on the chosen $s$ is implicit.

For fixed $W>\Wloc$, the slowest growth with observation time is approached by taking $s$ as large as the strict condition $r_s(W)<1$ permits. Denoting the supremum of such $s$ by $s_{+}(W)=\sup\{s\in(0,1):r_s(W)<1\}$, the infimum of the admissible time exponents is $1-s_{+}(W)$. This infimum is not attained with a finite decay length: since $r_s(W)$ diverges as $s\to1$, there must be an $s_{+}(W)<1$ where $r_{s_{+}(W)}(W)=1$ and the decay length diverges. As $W\searrow\Wloc$, $s_{+}(W)$ approaches the value $s_*\approx0.73$ that minimizes $(\sqrt{2\pi e}\,C_s)^{1/s}$ in Eq.~\eqref{eq:sstar} and locates $\Wloc$, giving the limiting exponent $1-s_*\approx0.27$. Increasing $W$ allows larger $s$ and hence smaller $\alpha$. The fixed choice $s=s_*$ remains admissible for every $W>\Wloc$.

\emph{Time windows and spatial confinement.} Positivity of $P_{\ell}(t)$ and $e^{-2} \le e^{-2t/T}$ for $0\le t\le T$ gives
\begin{equation}
    \frac{1}{T}\int_0^T\!\d t \, \E P_{\ell}(t)
    \le\frac{e^2}{2} \, \E\bar{P}_{\ell}(T),
\end{equation}
so the same spatial estimate holds for a sharp time window, up to a constant. Summing Eq.~\eqref{eq:dynamical_bound_detailed} over $\ell\ge L$ bounds the averaged probability of finding the wall beyond $L$, or equivalently the averaged probability $\langle P^\pm_L\rangle$ that spin $L$ is nonzero. For fixed $W$ and an admissible $s$, this tail is bounded by a constant times $(N+1)\,(WT)^{1-s}\,(L+2)^{s/2}\,e^{-L/\xi_s}\!$. It becomes small on a distance scale growing logarithmically with $N$ and $T$, with logarithmic corrections from the algebraic factor in $L$. This is an averaged occupation bound, not a bound on the maximum excursion at all times. 
Bounds uniform in time may be obtainable by adapting the eigenfunction-correlator method~\cite{AizenmanWarzelBook} to correlated disorder, which is a task considerably more involved.

\hypertarget{end:C}{\textit{C. Transfer recursion.}}---Here we discuss how the path sum $\Phi_{\ell}(s)$ can be efficiently computed via the recursion in Eq.~\eqref{eq:qtransfer}. The factorization in Eq.~\eqref{eq:factorization}, applied to $H$ without the auxiliary shift ($h_0=0$), gives $|G(0,\sigma;E)|^{s}=\prod_{j=0}^{\ell}|\Gamma_{\!\sigma^{(j)}}(E)|^{s}$, with $E$ short for $E+i0^{+}$. By the cavity relation $\Gamma_{\!\sigma^{(j)}}(z)=\cG_{j}\big(z-V(\sigma^{(j)})\big)$, each factor depends on its vertex only through the generation and the potential. We therefore group the branches by the potential of their endpoint, $\varphi_j(V) = \sum_{|\sigma|=j}\delta\big(V-V(\sigma)\big)\,|G(0,\sigma;E)|^{s}\!$, so that $\Phi_{\ell} = \int\!\d V \, \varphi_{\ell}(V)$ and $\varphi_0(V) = |\cG_0(E)|^{s}\,\delta(V)$. A vertex $\sigma^{(j)}$ of generation $j$ with potential $V(\sigma^{(j)})$ then fulfills
$|G(0,\sigma^{(j)};E)|^s=|G(0,\sigma^{(j-1)};E)|^s\,\big|\mathcal{G}_j\big(E-V(\sigma^{(j)})\big)\big|^s$. Multiplying both sides by $\delta(V-V(\sigma^{(j)}))$ and summing over all vertices in the same generation leads to the recursion in  Eq.~\eqref{eq:qtransfer}: the left side becomes $\varphi_j(V)$; the right side acquires an overall factor $|\cG_j(E-V)|^{s}\!$, and since the vertex descends from a parent at $V-h_j$ or $V+h_j$, the sum becomes $\varphi_{j-1}(V-h_j)+\varphi_{j-1}(V+h_j)$. In the local energy $u=E-V$ of a vertex, $\tilde{\varphi}_j(u)=\varphi_j(E-u)$ obeys
\begin{equation}
    \tilde{\varphi}_j(u) = \big|\cG_j(u)\big|^{s} \, \big[\tilde{\varphi}_{j-1}(u+h_j) + \tilde{\varphi}_{j-1}(u-h_j)\big],
    \label{eq:qtransfer_u}
\end{equation}
with $\tilde{\varphi}_0(u)=|\cG_0(u)|^{s}\,\delta(u-E)$ and $\Phi_{\ell} = \int\!\d u \, \tilde{\varphi}_{\ell}(u)$. The transfer operator in Eq.~\eqref{eq:qtransfer_u} contains only the functions $\cG_j$ and the fields, so they are the same at every energy, and $E$ only fixes where the walk starts. As the walk spreads over all energies within a few generations, it forgets its starting point, and $\gamma(s)$ should not depend on~$E$.
 
\emph{Numerics.} The recursion Eq.~\eqref{eq:cavity} is evaluated on a uniform grid $E\in[-E_{\max},E_{\max}]$ with spacing $\delta E$, using linear interpolation for the shifted arguments $z\mp h_{\ell+1}$ and the free asymptote $-1/z$ outside the grid. The path sum of Eq.~\eqref{eq:qtransfer} is propagated on the same grid with $\varphi_j$ normalized at every generation, for $N=400$, $E_{\max}=25$, and $\delta E=\eta/4$; reducing $E_{\max}$ to $15$ or increasing $\delta E$ to $4\eta$ moves the sign changes below by less than $0.005$. We read off $\gamma(s)$ from a fit over generations $100$ to $350$, which excludes the vicinity of the root and the last $50$ generations near the leaves; we find it to be self-averaging, with sample-to-sample fluctuations of order $10^{-2}$. The broadening cuts off the integrable singularities of $|\cG_j|^{s}$ at the poles of $\cG_j$ and therefore shifts the sign change of $\gamma(s=0.999)$ to smaller $W$: it lies at $W_c(\eta)=5.91$, $6.40$, $6.71$, and $6.90$ for $\eta=10^{-2}\!,\,\ldots,10^{-5}$. Fitting $\gamma(\eta)=\gamma_0-A\eta^p$ at each $W$ places the sign change of $\gamma_0$ at $\Wc=\Wcval\pm0.09$, which we quote; fitting $W_c(\eta)=\Wo-A\eta^p$ instead gives $\Wo=7.21\pm0.11$. These uncertainties are fit errors for the power-law form and do not include the systematic uncertainty from the choice of extrapolation. The fractional power $s<1$ makes the poles integrable, which is what allows the sum over $2^{\ell}$ discrete branches to be replaced by a density in $V$.
 
The typical local density of states of Fig.~\ref{fig:dynamics}b is evaluated with $\delta E=\eta/5$ and $E_{\max}=2\sqrt2+4W$, at $E=1$ because the band center is special: by chiral symmetry the eigenstate that localizes on the root has energy exactly $E=0$---for the three-site cluster of the root and its children the zero mode $(h_1,-1,1)/\sqrt{h_1^2+2}$ carries the root weight $h_1^2/(h_1^2+2)$---so that $\rho(0)\to w_0/(\pi\eta)$ in the localized phase. Evaluating the same runs at $E=0$ gives essentially the mirror image of Fig.~\ref{fig:dynamics}b: $\rho_{\rm typ}(0)$ is independent of $\eta$ at weak disorder and grows as $1/\eta$ at strong disorder, where $\pi\eta\rho(0)$ approaches a finite, $\eta$-independent zero-mode weight, with the same broad crossover in between as for $\rho_{\rm typ}(1)$.
The survival probability of Fig.~\ref{fig:dynamics}a is obtained from separate runs ($N=400$, $10^3$ realizations, $\eta=10^{-4}$, $E_{\max}=25$, $\delta E=\eta/4$): $A(t) = \int\!\d E \, \rho(E)e^{-iEt}$ is evaluated on the grid, and the broadening is removed exactly through $A_\eta(t) = A(t)e^{-\eta t}$. Since this amplifies the numerical errors from the finite grid and energy window by $e^{\eta t}$, the result is reliable for $t\lesssim1/\eta$, and the times $t\le10^3$ shown at $\eta=10^{-4}$ lie well within this range. Within the plotted range, $P_0(t)$ agrees with the result at $\eta=10^{-3}$ to within $0.2\%$ and changes by less than $0.02\%$ in relative terms when $\delta E$ is doubled or $E_{\max}$ is reduced to $15$.

\begin{figure}[t]
    \includegraphics[width=\linewidth]{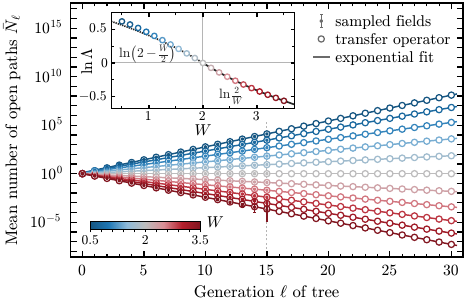}
    
    \caption{\label{fig:num_paths}%
    \textbf{Resonant branches.} Mean number $\bar{N}_{\ell}$ of branches along which the potential of Eq.~\eqref{eq:potential} stays within the resonance window $|V|\le1$ up to generation $\ell$, for disorder strengths $W=0.5,\dots,3.5$. Dots: direct count on sampled fields ($10^5$ realizations for $W<2$, $10^7$ for $W\ge2$; $3\sigma$ error bars). Circles: transfer operator, Eq.~\eqref{eq:transfer}, evaluated on a grid of $2^{32}$ points using quantics tensor networks~\cite{Waintal2026}. Lines: exponential fits for $\ell\ge10$. The growth turns into decay exactly at $\Wcl=2$. Inset: fitted rates compared with $\ln\Lambda=\ln(2/W)$ for $W\ge 2$ (solid) and the bound $\ln(2-W/2)$ for $W<2$ (dotted).}
\end{figure}

\hypertarget{end:D}{\textit{D. Resonant branches.}}---We count the branches along which the potential stays within the resonance window $|V|\le1$ set by the unit hopping. Let $n_{\ell}(V)$ be the density of such branches at potential $V$ in generation $\ell$, with $n_0(V)=\delta(V)$. A vertex at potential $V'$ has two children at $V'\pm h_{\ell+1}$, of which those inside the window survive, so $n_{\ell+1}(V)=\Theta(1-|V|)[n_{\ell}(V-h_{\ell+1})+n_{\ell}(V+h_{\ell+1})]$. Averaging over $h_{\ell+1}$ uniform on $[-W,W]$ gives the transfer operator
\begin{equation}
    (\Tw n_{\ell})(V) = \frac{1}{W} \int_{-1}^{1}\!\d V' \, \Theta\big(W-|V-V'|\big) \, n_{\ell}(V'),
    \label{eq:transfer}
\end{equation}
and since the fields of different generations are independent, the mean number of resonant branches is $\bar{N}_{\ell} = \int_{-1}^1\!\d V \, (\Tw^{\ell} n_0) \sim \Lambda(W)^{\ell}$, with $\Lambda$ the largest eigenvalue of the symmetric positive operator $\Tw$. For $W\ge2$, $\Theta(W-|V-V'|)=1$ for all $V,V'$ in the window, so $\Tw$ maps every $n_{\ell}$ to the constant $W^{-1} \int_{-1}^{1}\!\d V' \, n_{\ell}(V')$ and has the single nonzero eigenvalue $\Lambda=2/W\le1$. For $W<2$, the expectation value of $\Tw$ with respect to the constant function $1/\sqrt{2}$ gives the lower bound $\Lambda\ge\frac{1}{2}\int_{-1}^{1}\!\d V' \, (\Tw 1)(V') = 2-W/2 > 1$, since the integral is $W^{-1}$ times the area $4-(2-W)^2$ of the band $|V-V'|\le W$ inside the square $[-1,1]^2$. The mean number of resonant branches thus grows exponentially for $W<\Wcl=2$ and decays beyond, as confirmed by the direct counts in Fig.~\ref{fig:num_paths}. A binary tree with independent potentials, uniform on $[-W, W]$ at each vertex, has the same threshold: each child is resonant independently with probability $p=\min(1,1/W)$, so the mean number of resonant branches grows by the factor $2p$ per generation. The mechanism differs, however: on that tree, all branches remain resonant for $W\le1$; in our model, the random walk of the potential leaves the window on almost all branches for any $W>0$, and only a fraction $\Lambda/2<1$ of the children of resonant vertices remains resonant on average.
\end{document}